\documentclass{ws-ijmpc}

\begin{document}

\markboth{G\'eza \'Odor}
{Self-organizing, two-temperature Ising model describing human segregation}

\catchline{}{}{}{}{}

\title{Self-organizing, two-temperature Ising model describing
human segregation}

\author{G\'eza \'Odor}

\address{Research Institute for Materials Science, P.O.Box 49,\\
Budapest, H-1525, Hungary\
odor@mfa.kfki.hu}

\maketitle

\begin{history}
\received{Day Month Year}
\revised{Day Month Year}
\end{history}

\begin{abstract}
A two-temperature Ising-Schelling model is introduced and studied for 
describing human segregation. The self-organized Ising model with
Glauber kinetics simulated by M\"uller et al. exhibits a phase 
transition between 
segregated and mixed phases mimicking the change of tolerance 
(local temperature) of individuals. The effect of external noise is 
considered here as a second temperature added to the decision of 
individuals who consider change of accommodation. A numerical 
evidence is presented for a discontinuous phase transition 
of the magnetization.
\end{abstract}


\section{Introduction}

The problem of human segregation is an important problem of society 
and politics even in the 21st century \cite{1}. Social sciences have been 
investigating the reasons and and nature of segregation for a long time.
Sociologist have introduced several models, one of them is the Schelling
model \cite{2}. From physicist point of view that model is a 3-state
voter-type non-equilibrium model (groups A,B and empty), with spin-exchange
dynamics at zero temperature ($T=0$) on a 2-dimensional square lattice. 
Although the model describes a segregation by a quench without
external reasons, unwanted frozen states may also occur.

Recently it was shown by computer simulations \cite{3} that there 
exists a simpler model, namely the Glauber-Ising model \cite{4,5}, which 
captures the essence of human segregation. Besides that, the usage of
the simple $T>0$ spin-flip dynamics makes it computationally easier and
one can avoid the frozen states of the Schelling model. In this model 
the temperature plays the role of a global tolerance; by varying it,
the model may or may not evolve into the ordered (segregated) state.

However a constant, global tolerance is rather artificial in a society,
it can vary from individual to individual and can change in time
as well. By introducing a local, time dependent temperature (tolerance),
with a feedback mechanism from the local neighborhood the model becomes 
more realistic. The results do not change too much \cite{6}, a
self-organization of the average temperature occurs. The parameters
of this model are the global rate of forgetting (of tolerance) 
and the response of local tolerances on the neighborhood. The sum of 
these local changes determine the local temperature. Hereafter this
self-organized segregation model will be called SO-Seg model.

Stepping further towards more realistic models, one can pose the question
what happens to this model if the decision of individuals are affected by 
an independent external noise as well. The external noise can be an 
artifact of a random environment, housing, moving situation, presence of
shopping centers ... etc. The external noise introduced here as a second
temperature, i.e. individuals are connected to a second heath bath.

Two-temperature two-state voter-type models have been investigated 
intensively recently and have become the prototypes of non-equilibrium 
models (for a review see \cite{7}). An important finding of these 
studies was the discovery of relevant factors affecting the phase 
transitions of models exhibiting $Z_2$ (up-down) symmetry \cite{8}. 
In particular models with general, isotropic spin-flip dynamics 
maintaining the $Z_2$ symmetry can be classified as two temperature 
models, where one temperature controls the bulk, the other the 
interface fluctuations.
The Ising model is a special case of these models, where both temperatures
are nonzero and the time-reversal symmetry drives the system into an
equilibrium state. The transition of these $Z_2$ symmetric, two-temperature
models has been found to be continuous, Ising type unless the bulk 
temperature is zero. In the latter case it is first order, voter 
model class type \cite{7}.
In this work I investigate the effect of a second temperature applied 
as an external, independent heat-bath to the spins of the SO-Seg model.

\section{The model}

The model is defined on 2-dimensional square lattice, with periodic boundary
conditions and Ising spins ($s_i=(1,-1)$) distributed initially randomly
(zero initial magnetization = no segregation). The kinetics follows 
a Glauber spin-flip (sequential) update with Heath Bath acceptance rate
(see Ref.~\refcite{9}), depending on the local temperatures. 
In the SO-Seg model each individual has four interacting nearest
neighbors and a randomized initial local temperature, with an average 
value $<T_1(0)> = 1.5$. This local temperature is lowered by $\delta
T_F$ at each update for modeling the loss of tolerance. 
This alone would just make a quench to $T_1\to 0$ with domain coarsening. 

To model people's awareness of the dangers of segregation they can 
increase their own temperature (tolerance) by $\delta T_a$, if all four 
neighbors of an individual belong to the same group as $s_i$. If all four
neighbors belong to the different group the local temperature is decreased
by the same amount.

Our external noise is described by a heat bath of a second temperature 
$T_2$ applied to the decisions. The actual spin-flip will be the 
logical OR of internal and external flip decisions. The magnetization 
$M(t)=<s_i(i)>$, the average number of 
like neighbors minus unlike neighbors $<N(t)>$,
and the average self-organizing internal temperature (tolerance) 
$<T_1(t)>$ is followed up to $t_{max}=2\times 10^5$ Monte Carlo 
sweeps (MCS) of the lattice.

\section{Simulation results}

The simulations were performed on $L=400$, $2000$ and $4000$ 
sized square lattices, up to $t_{max}=2\times 10^5$ MCS, with 
cooling rates: $\delta T_F=0.01,0.02$ and tolerance steps:
$\delta T_a=0.003, 0.002$.
The phase transition of the SO-Seg model is at $\delta T_a=0.0029$
\cite{5}, so $\delta T_a=0.003$ corresponds to super-critical, 
$\delta T_a=0.002$ to sub-critical situations.

\begin{figure}[hp]
\centerline{\psfig{file=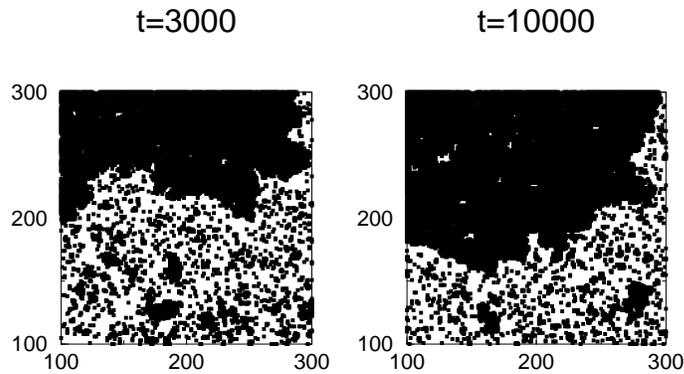,width=9cm}}
\vspace*{8pt}
\caption{Clusters survive small external noise: $T_2=0.5$.
\label{fig0}}
\end{figure}

\begin{figure}[hp]
\centerline{\psfig{file=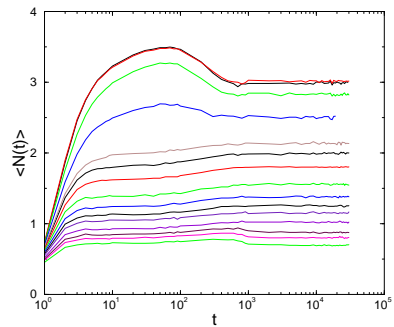,width=5.1cm}}
\vspace*{2pt}
\caption{Average composition as the function of time and 
external temperature $T_2=0,1,2...10$ (top to bottom)
\label{fign}}
\end{figure}

\begin{figure}
\centerline{\psfig{file=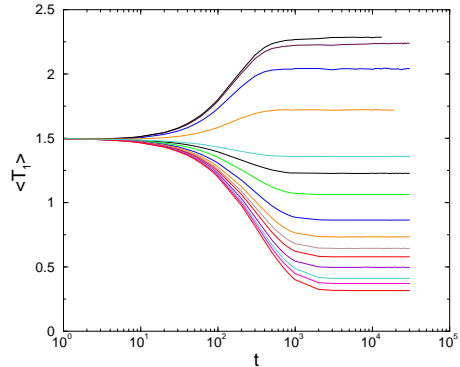,width=6cm}}
\vspace*{2pt}
\caption{The average tolerance of individuals as the function of
time and external temperature $T_2=0,1,2...10$ (top to bottom)
\label{figT}}
\end{figure}

\begin{figure}
\centerline{\psfig{file=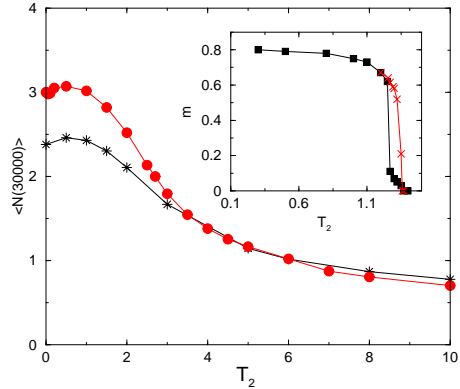,width=6cm}}
\vspace*{2pt}
\caption{Average composition as the function of external temperature
($T_2$). Bullets (higher data on the left) correspond to $\delta T_a=0.003$
forgetting rate, crosses to $\delta T_a=0.002$ forgetting rate.
Inset: The magnetization density for $\delta T_a=0.003$ shows
a first order transition.
\label{figninf}}
\end{figure}

As Figure (\ref{fig0}) shows the inclusion of a small second 
temperature $T_2 < \sim 1$ does not change the composition of neighbors
and $T_1$ in the steady state (as $t\to\infty$).
The same can be seen by plotting $N(t)$ (see Fig.(\ref{fign})) 
and the average internal temperature of individuals 
(Fig.(\ref{figT})). 
For stronger external noise the domains are destroyed, but the 
average tolerance goes to zero too.
This means that the unsegregated state can be maintained with
the help of strong external noise, without worrying about
people's local tolerance.
The same analysis for $\delta T_a=0.002$ resulted in similar trends
in  $N(t=30000)$ (see Fig.\ref{figninf}) and in the average
tolerance $T_1(t=30000)$.
Note that one observe even a weak increase in the asymptotic values
for $T_2<1$. Running the simulations on larger sizes there were no
change in this, excluding the possibility of finite size errors.

By increasing the second temperature $T_2$ the transitions of 
$N(\infty)$ (see Fig.(\ref{figninf})) and $T_1(\infty)$ 
are very smeared. The magnetization density
$m = M(\infty) / L^2$ on the other hand shows a sharp fall, 
indicating a first order phase transition at $T_2^*=0.135(2)$.
A fitting attempt using the form $m = A (T_2^* -T_2)^{\beta}$ did
not result in agreement with the 2d Ising class continuous
phase transition behavior, characterized by $\beta = 1/8 $ \cite{7}.
Furthermore a hysteresis cycle can also be found by starting
the simulations with different initial conditions 
(ordered vs. disordered), which is a clear hallmark of a 
first order phase transition. 
This is in contrast with the results for two-temperature 
$Z_2$ symmetric models exhibiting Ising transition in 2d 
(the bulk noise in nonzero of course). 
One may understand the discontinuous transition here by realizing that 
this model is effectively a coupled system: Ising + temperate field.
The temperatures can increase if at least four neighbors are in the
same state and decrease without condition. In the language of reaction
diffusion systems it is a quadruple model, where at least for
neighbors are needed for a creation but the removal is spontaneous.
It is well known that in 2d such a quadruple model exhibits 
a first order transition \cite{7}.

Simulations with different initial temperatures ($T_1(0)=2.5$), 
cooling rates ($T_F=0.2$) and $\delta T_a$ 
have not resulted in changes in the transition point $T_2^*$. 

\section{Conclusion}
A two-temperature, self-organized Ising-Schelling model has been
introduced and investigated by numerical simulations. A low second 
temperature, which represents the external noise to the decision 
of individuals for moving does not change the segregation
behavior of the model. A temperature bigger than $T_2=0.135(2)$
randomizes the segregated, ordered domains and results in 
low average tolerance of individuals. While the self-organized 
tolerance and the average composition of the steady state show 
continuous variation on $T_2$ the magnetization exhibits a first
order phase transition. The threshold did not show considerable 
dependence on the model parameters.

\section*{Acknowledgments}
The author thanks D. Stauffer for motivating and discussing
his study and providing the code of Ref. \cite{6} for simulations.
Support from the Hungarian Research Fund OTKA (Grant No. T-046129) 
during this study is gratefully acknowledged.



\end{document}